\documentclass[12pt]{iopart}

\def\sqrtsNN{\mbox{$\sqrt{s_\mathrm{NN}}$}}
\usepackage{graphicx}
\begin{document}

\title[Non-photonic electron-hadron correlations at STAR]{Non-photonic electron-hadron correlations at STAR}

\author{Gang Wang (for the STAR\footnote{For the full list of STAR authors and acknowledgements, see appendix `Collaborations' in this volume.} Collaboration)}

\address{University of California,
Los Angeles, California 90095, USA}
\ead{gwang@physics.ucla.edu}

\begin{abstract}
We present STAR's measurements of azimuthal correlations between
non-photonic electrons and charged hadrons in p+p, Cu+Cu and Au+Au collisions at $\sqrtsNN = 200$ GeV. 
In p+p collisions, from the non-photonic e-h correlation we have extracted the relative B meson semi-leptonic
decay contributions to non-photonic electrons up to electron $p_t \sim 9$ GeV/$c$.
In central Cu+Cu and Au+Au collisions where a dense medium is created, we find that the
e-h correlation on the away side of the trigger non-photonic electron has been modified in comparison with 
the expectation from PYTHIA simulations. 
\end{abstract}

%\maketitle

\section{Motivation}
In central Au+Au collisions at RHIC, electrons from heavy quark (charm and bottom) decays
have been observed to be suppressed to a similar level as light hadrons~\cite{RHIC_RAA1,RHIC_RAA2}.
It is of great interest to separate non-photonic electron energy loss into
the contributions from charm and bottom quarks.
Since the near-side e-h azimuthal correlation from B decays is 
much wider than that from D decays for the same electron $p_t$,
STAR's previous study~\cite{Xiaoyan} compared the experimental correlation 
results in p+p collisions with PYTHIA simulations,
and found a substantial B contribution to non-photonic electrons up to electron $p_t \sim 6$ GeV/$c$.
In this work, we extend this measurement to $p_t \sim 9$ GeV/$c$.

When high-$p_t$ partons lose a significant amount of energy through
the dense QCD medium created in central Cu+Cu or Au+Au collisions,
their azimuthal correlations with low-$p_t$ hadrons are modified accordingly,
showing a broad or even double-peak structure on the away-side di-hadron correlation~\cite{STAR_corre,Horner}.
Non-photonic electrons represent well the directions of the mother D or B mesons
when electron $p_t >$ 3 GeV/$c$, and in the opposite direction we expect another 
heavy quark traversing the medium leaving an imprint on the away-side e-h correlation.
This study of the correlated hadron azimuthal distribution from heavy quark energy loss in the dense QCD
medium will provide insight on the mechanism responsible for the correlation pattern and
the flavor dependence.
Cu+Cu and Au+Au collisions have very different system sizes, and enable us to have a systematic study
on the system-size dependence of the non-photonic e-h correlation.

\section{Data sets and the study}
This study is based on STAR events with high-tower triggers at 200 GeV: 
0.7 million p+p events in RUN VI, 1 million $0 - 20\%$ most central Cu+Cu events in RUN V,
and 0.2 million $0 - 20\%$ Au+Au events in RUN VII. 
Tracks of charged particles are reconstructed by the STAR TPC~\cite{TPC}.
To enhance the statistics for high $p_t$ electrons, the high-tower trigger requires that 
at least one track is projected into the STAR BEMC and its SMD~\cite{EMC},
with the energy deposition above a threshold in a single tower of the BEMC.
The high-tower thresholds are $5.4$ GeV (p+p), $3.75$ GeV (Cu+Cu) and $5.5$ GeV (Au+Au).
The pseudorapidity ($\eta$) coverage in this study is $0<\eta<0.7$ for Cu+Cu,
and $|\eta|<0.7$ for p+p and Au+Au collisions, to avoid the large photon conversion background for $|\eta|>0.7$.
The centrality definition used is the same as in Ref.~\cite{Centrality}. 

Electron identification was carried out by combining the ionization energy loss in the TPC
with the energy deposition in the BEMC and the shower profile in the SMD.
(See details in \cite{RHIC_RAA1} and \cite{Weijiang}.)
The dominant photonic electron background is from photon conversions and $\pi^0$, $\eta$
Dalitz decays, where electron pairs have small invariant masses.
We pair up the electron candidates with tracks passing a very loose cut on $d$E$/dx$ around the electron
band, and examine the $2$-$D$ invariant mass distribution of electron pairs with opposite signs ($OppSign$),
by ignoring the opening angle $\Delta \phi$ to eliminate the tracking resolution effects~\cite{Weijiang}.
The $OppSign$ electron sample contains the true photonic background as well as
the combinatorial background estimated by the same-sign pairs ($SameSign$).
A cut of mass $<0.1$ GeV/$c^2$ rejects most photonic background. 

To study the angular correlation of non-photonic electrons and charged hadrons, 
we start with the semi-inclusive electron sample ($semi$-$inc$),
which is everything in the inclusive electron sample %($inc$) 
except for the $OppSign$ with the mass cut. 
The latter term contains the reconstructed photonic electrons %($reco$-$pho$)
and the $combinatorics$, or the $SameSign$ with the mass cut.
%On the other hand, $reco$-$pho$ is the photonic electron sample ($pho$) 
%without the not-reconstructed photonic electrons ($not$-$reco$-$pho$),
%and $inc = pho + non$-$pho$.
%Thus we have $semi{\textrm-}inc = inc - (reco$-$pho + combinatorics) = non$-$pho + not$-$reco$-$pho - combinatorics$.
The correlation signal is then obtained by the equation 
$\Delta \phi_{non{\textrm-}pho} = \Delta \phi_{semi{\textrm-}inc} - \Delta \phi_{not{\textrm-}reco{\textrm-}pho} + \Delta \phi_{combinatorics}$.
Here $\Delta \phi_{not{\textrm-}reco{\textrm-}pho} = (1/\epsilon - 1) \Delta \phi_{rec{\textrm-}pho{\textrm-}no{\textrm-}partner}$,
where $\epsilon$ is the photonic electron reconstruction efficiency, and 
$\Delta \phi_{rec{\textrm-}pho{\textrm-}no{\textrm-}partner}$ denotes the reconstructed photonic electron
azimuthal correlations with charged hadrons after removing the photonic partners. 
The efficiency $\epsilon$ was estimated from simulations to be $\sim 70\%$ for p+p,
$\sim 66.5\%$ for Cu+Cu, and $\sim 80\%$ for Au+Au.
The details on the approach via $semi$-$inc$ can be found in \cite{Xiaoyan}.

\section{Results}

Fig.~1 shows the B contribution to non-photonic electrons as a function of electron $p_t$
for 200 GeV p+p collisions, where the open stars represent the results from RUN V data~\cite{Xiaoyan}
and the full stars, RUN VI data. Each point in Fig.~1 was obtained from the azimuthal
correlation between the non-photonic electron with the corresponding $p_t$ and the charged hadrons
with $p_t > 0.3$ GeV/$c$, after we fit the near side correlation with PYTHIA simulations. 
(See the fitting recipe in \cite{Xiaoyan}.)
The systematical errors come from photonic electron reconstruction efficiency,
different fitting ranges and different fitting functions. The solid curves show the range of relative bottom contribution 
from recent pQCD calculations (FONLL)~\cite{fonll}\footnote{FONLL calculations with CTEQ6M, $m_c = 1.5$ GeV/$c^2$,
$m_b = 5$ GeV/$c^2$ and $\mu_{R,F} = m_T$ (for the central value of the range).}.
Preliminary STAR data are consistent with the FONLL calculations,
and show almost the same B and D contributions to non-photonic electrons for $p_t > 6$ GeV/$c$.
Together with the observed suppresion of non-photonic electrons in central Au+Au collisions,
the measured $e_B / (e_B+e_D)$ ratios imply that bottom may be suppressed in central
Au+Au collisions at RHIC, assuming the initial ratio is similar in Au+Au as in p+p.

%\begin{figure}
%\includegraphics[width=1.0\textwidth]{B_ratio_all.eps}
%\caption{B contribution to non-photonic electrons as a function of electron $p_t$
%for 200GeV p+p collisions.
%The error bars are statistical, and the error brackets are systematical.}
%\label{fig:pp}
%\end{figure}

\begin{figure}[!btp]
\begin{minipage}[r]{0.7\textwidth}
\center
\includegraphics[width=\textwidth]{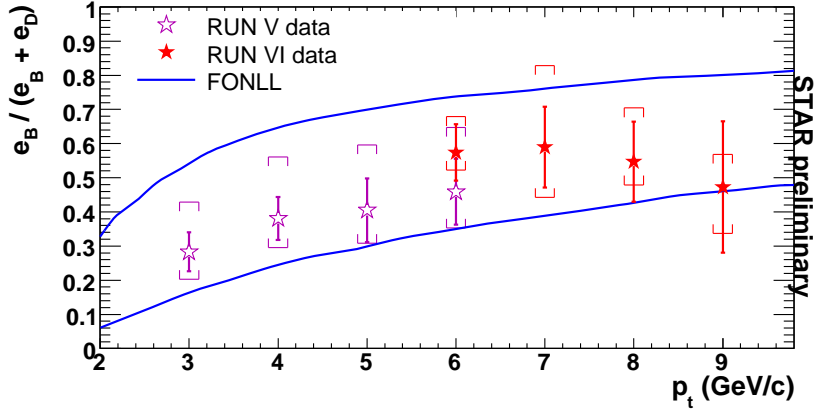}
\label{fig:pp}
\end{minipage}
\begin{minipage}[c]{0.29\textwidth}
  \caption{B contribution to non-photonic electrons as a function of electron $p_t$ for 200GeV p+p collisions. 
  The error bars are statistical, and the error brackets are systematical.}
\end{minipage}
\vspace{-0.5cm}
\end{figure}

\begin{figure}
\hspace{1cm}
\includegraphics[width=0.85\textwidth]{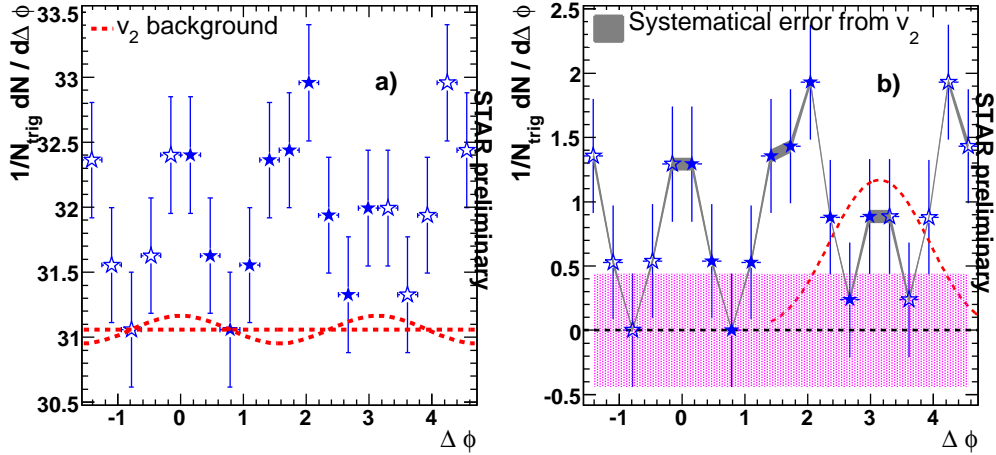}
\vspace{-0.5cm}
\caption{Non-photonic e-h correlation in Cu+Cu collisions at 200 GeV. Panel a) shows the correlation
before $v_2$ background subtraction, and panel b), after the subtraction, with a dashed fitting curve
from PYTHIA expectations on the away side.
The error bars are statistical, and the error band around zero shows the systematical uncertainty from ZYAM.}
\label{fig:CuCu}
\end{figure}

\begin{figure}
\hspace{1cm}
\includegraphics[width=0.85\textwidth]{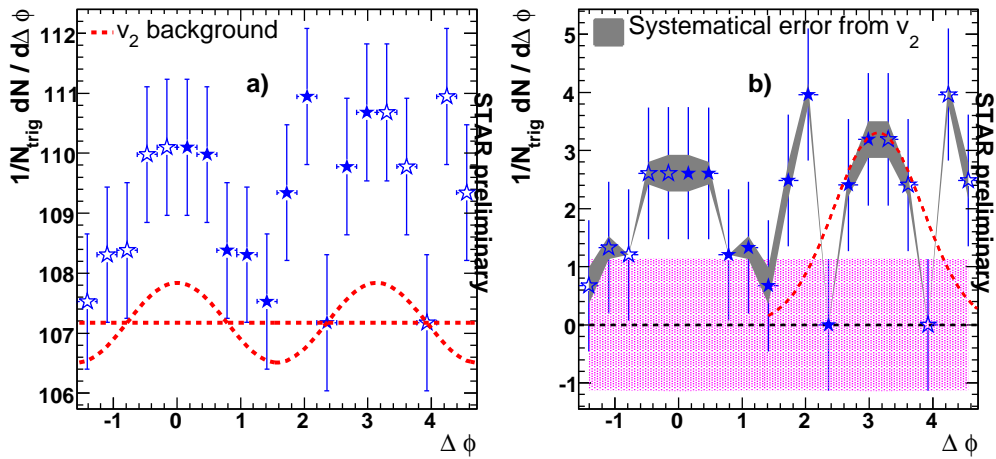}
\vspace{-0.5cm}
\caption{Non-photonic e-h correlation in Au+Au collisions at 200 GeV. Panel a) shows the correlation
before $v_2$ background subtraction, and panel b), after the subtraction, with a dashed fitting curve
from PYTHIA expectations on the away side.
The error bars are statistical, and the error band around zero shows the systematical uncertainty from ZYAM.}
\label{fig:AuAu}
\end{figure}

Fig.~\ref{fig:CuCu} and Fig.~\ref{fig:AuAu} show non-photonic e-h correlations at 200 GeV 
Cu+Cu and Au+Au collisions, respectively.
The raw azimuthal correlations are plotted in panels a), with elliptic flow ($v_2$)~\cite{Methods} background curves
determined with the Zero Yield at Minimum (ZYAM) approach~\cite{STAR_corre}.
The lower limits of $v_2$ are zero, and the upper limits are taken to be
$60\%$ (for Cu+Cu) or $80\%$ (for Au+Au) of charged hadron $v_2$ measured for the same 
$p_t$ ranges as the trigger electron ($3$ GeV/$c < p_t < 6$ GeV/$c$) and the associated
hadron ($0.15$ GeV/$c < p_t < 0.5$ GeV/$c$ for Cu+Cu, and $0.15$ GeV/$c < p_t < 1$ GeV/$c$ for Au+Au).
To enhance the statistics, the measured correlations are folded into $[0, \pi]$, 
and the data points beyond are reflections.
The azimuthal correlations after the subtraction of $v_2$ background are shown in panels b).
Despite large statistical errors, clear correlation structures are present for both Cu+Cu and Au+Au.
On the near side, there is one single peak representing the heavy quark fragmentation, and
possible interactions with the medium.
On the away side, instead of one peak around $\pi$ as in p+p collisions, the correlation functions
are modified to be a broad or possible double-peak structure.
A single peak structure expected from PYTHIA calculations can not describe the measured away side correlations.
A PYTHIA fit to the away side structure yields values of $\chi^2 / ndf$ to be $54.23 / 10$ (Cu+Cu) and $25.68 / 10$ (Au+Au).
The away side broadness in both Cu+Cu and Au+Au is similar to the di-hadron correlation in Au+Au~\cite{Horner},
and probably indicates heavy quark interaction with the dense medium.

\end{document}